\PassOptionsToPackage{bookmarks=false}{hyperref}
\documentclass[sigconf,authorversion,nonacm]{acmart}

\usepackage{amsmath}
\usepackage{amsthm}
\usepackage{booktabs}
\usepackage{comment}
\usepackage{courier}
\usepackage{helvet}
\usepackage[utf8]{inputenc}
\usepackage{mathtools}
\usepackage{soul}
\usepackage{tabularx}
\usepackage{times}
\usepackage{url}

\DeclareMathOperator*{\argmax}{arg\,max}

\title{Fairness and Sequential Decision Making: \\ Limits, Lessons, and Opportunities}

\author{Samer B. Nashed}
\affiliation{%
 \institution{University of Massachusetts Amherst}
 \country{USA}}
\email{snashed@cs.umass.edu}
 
\author{Justin Svegliato}
\affiliation{%
 \institution{University of California, Berkeley}
 \country{USA}}
\email{jsvegliato@berkeley.edu}
 
\author{Su Lin Blodgett}
\affiliation{%
 \institution{Microsoft Research}
 \country{Canada}}
\email{sulin.blodgett@microsoft.com}

\begin{document}

\begin{abstract}
    As automated decision making and decision assistance systems become common in everyday life, research on the prevention or mitigation of potential harms that arise from decisions made by these systems has proliferated. However, various research communities have independently conceptualized these harms, envisioned potential applications, and proposed interventions. The result is a somewhat fractured landscape of literature focused generally on ensuring decision-making algorithms ``do the right thing''. In this paper, we compare and discuss work across two major subsets of this literature: algorithmic fairness, which focuses primarily on predictive systems, and ethical decision making, which focuses primarily on sequential decision making and planning. We explore how each of these settings has articulated its normative concerns, the viability of different techniques for these different settings, and how ideas from each setting may have utility for the other.
\end{abstract}

\maketitle

\section{Introduction}

The social and ethical implications of different technologies have long been the object of study for scholars outside of computer science, and recently many computer scientists have taken up this broader agenda under a variety of names. In particular, two largely independent communities have evolved from established fields of computer science. The study of algorithmic fairness that has emerged at the FAccT conference and its predecessors is heavily influenced by the field of machine learning and focuses on predictive systems, while the study of ethical decision making\footnote{The term ``ethical decision making'' has (unsurprisingly) been used to describe a variety of research, including symbolic planning and system verification, but here we use it to refer to work focusing on ethical concerns arising from sequential decision making systems.} has attracted primarily researchers from classical artificial intelligence and focuses on sequential decision making. Nominally, these groups have similar goals: to produce predictive or decision-making systems that ``do the right thing''. However, many key ideas from ethical decision-making have not yet percolated into the fairness literature, and likewise many important concepts and approaches developed for fair prediction are not yet common in ethical decision making. This paper is an effort to bridge this gap.

Unlike predictive systems, which consider decisions independently and one at a time (known as myopic decision making), sequential decision-making systems consider sequences of potential actions, allowing them to evaluate the long-term effects of taking a particular set of actions. Many real-world problems, such as autonomous driving, power grid management, wildfire fighting, military engagement, disaster relief, and inventory logistics, both fundamentally affect people's safety and access to resources and require sequential reasoning as they cannot be solved adequately via myopic decision making. However, although problems such as autonomous driving sometimes motivate the fairness literature~\cite{fish2016confidence,stilgoe2021we,hannan2021gets,finocchiaro2021bridging,khakurel2022exploring,londono2022fairness,tao2022ruler}, fairness conceptualizations and methods have largely been developed for predictive rather than sequential decision-making systems. Moreover, despite the fairness literature's acknowledgement of the long-term effects and sequential nature of many high-stakes decisions \cite{ensign2018runaway,chouldechova2018case,liu2018delayed,hashimoto2018fairness,hu2019disparate,elzayn2019fair,milli2019social,rodolfa2020case,bountouridis2019siren,noriega2020algorithmic,donahue2021better,reader2022models,schwobel2022long,manshadi2021fair,gupta2021individual}, including education and college admissions~\cite{heidari2021allocating,mouzannar2019fair,almuzaini2022abcinml}, recidivism risk prediction \cite{d2020fairness,lum2020impact}, predictive policing \cite{chapman2022data}, child and homeless welfare \cite{du2022data,rahmattalabi2022learning}, clinical trials \cite{chien2022multi}, and hiring \cite{liu2020disparate,blum2022multi}, work on these settings rarely engages problem formulations or approaches developed for sequential decision making, or efforts to conceptualize and address ethical concerns emerging from the ethical decision making literature.

Our paper makes the following contributions. We begin by introducing a foundational and widely-used sequential decision-making model, the Markov decision process (MDP), from which many special-case models are derived. We cover problem formulation, solution methods, and key assumptions and properties (\S\ref{sec:background}). We then examine how ethical concerns have been conceptualized within the ethical decision-making and fairness literatures (\S\ref{sec:conceptualizations}), examine the sequential decision-making model pipeline (\S\ref{sec:pipelines}), introduce some of the measurements (\S\ref{sec:measurement}) and mitigations (\S\ref{sec:solutions}) common in the ethical decision-making literature, and discuss some current challenges and state-of-the-art techniques for ethical decision making.

Throughout, we offer observations following three general themes. First, we draw comparisons between conceptualizations, measurements, and mitigations proposed in the fairness and ethical decision making literatures to highlight where insights and methods from fairness may or may not be appropriate for ethical decision making. Second, inspired by the fairness literature's analyses of machine learning pipelines, we draw attention to aspects of sequential decision-making pipelines that represent open opportunities for future analysis. Finally, we highlight some problem formulations and techniques developed for ethical decision making that may offer advantages for fairness research.

\section{Background on Sequential Decision Making} \label{sec:background}

A \textbf{Markov decision process (MDP)} is a general sequential decision-making model\footnote{MDPs and their variants occupy the vast majority of the AI and planning literature that uses the term ``sequential decision making''.} that enables an agent\footnote{We use the terms ``agent'' (the preferred term in classical AI research) and ``system'' (a more general, catchall term) interchangeably to describe collections of processes which can take actions in the world, such as a robot. We use the term ``model'' to describe a decision-making or predictive model specifically, removed from the larger system in which it operates.} to make a sequence of decisions in fully observable, stochastic environments~\cite{bellman1966dynamic} and has been used in many decision-making problems, such as search and rescue~\cite{goodrich2008supporting,pineda2015continual}, extraterrestrial exploration~\cite{mustard2013mars,gao2017review}, and autonomous driving~\cite{wray2016hierarchical,wray2017online,basich2020learning}. An MDP describes a decision-making problem using four attributes: (1) a set of \textbf{states} that represent different possible scenarios, (2) a set of \textbf{actions} that can be performed by the agent, (3) a \textbf{transition function} that gives the probability of reaching a given state when the agent performs a particular action in its current state, and (4) a \textbf{reward function} that gives the immediate utility of performing a particular action in its current state. At each time step the agent performs an action in a state, receives a reward based on the reward function, and transitions to a successor state based on the transition function. MDPs satisfy a key property, called the \textbf{Markov property}, that holds that the outcome of any action only depends on the current state. That is, the agent's prior states and actions do not matter. The solution to an MDP is the \textbf{optimal policy}, the mapping from states actions that maximizes the value function. The \textbf{value function} is defined over all states and represents the expected cumulative reward the agent would earn if it executed the optimal policy from each state.

{\bf Formal Definition: } An MDP is a tuple, $\langle S, A, T, R \rangle$, where: $S$ is a finite set of states; $A$ is a finite set of actions; $T(s, a, s')$ is a transition function that represents the probability of reaching state $s'$ after performing action $a$ in state $s$; and $R(s, a)$ is a reward function that represents the immediate reward gained by performing action $a$ in state $s$. At each time step, the agent performs an action $a$ in a state $s$, experiences reward $R(s, a)$, and transitions to a successor state $s'$ with probability $T(s, a, s')$. The agent either repeats these steps forever (infinite horizon) or until a deadline (finite horizon).

A solution to an MDP is a policy $\pi: S \rightarrow A$, where $\pi(s) = a$ indicates that the agent should perform action $a$ when in state $s$. For a given policy $\pi(s)$, its value function $V^\pi(s)$ describes the value of each state $s$ with respect to the policy $\pi(s)$. In particular, the value function $V^\pi(s)$ describes the expected cumulative reward that the agent would earn starting in state $s$ and executing policy $\pi(s)$, until reaching the horizon:

\[ V^\pi(s) = R(s, \pi(s)) + \gamma \sum_{s' \in S} T(s, \pi(s), s') V(s'). \] Typically, the expected cumulative reward is discounted to balance the value of immediate rewards with the value of future rewards: that is, the discount factor is often $\gamma \in [0, 1)$ in infinite horizon MDPs and $\gamma = 1$ in finite horizon MDPs. Along with balancing rewards gained in the present and rewards gained in the future, a discount factor $\gamma <1 $ provides guarantees that the value function of an infinite horizon MDP converges to finite values. The goal of the agent is to find the optimal policy $\pi^*(s)$ that maximizes the value---the expected cumulative reward---of each state $s$ until reaching the horizon:
\[ V^*(s) = \max_{a \in A} \big[ R(s, a) + \gamma \sum_{s' \in S} T(s, a, s') V^*(s') \big]. \]
Finally, given the optimal value function $V^*(s)$, the optimal policy $\pi^*(s)$ can be calculated in the following way:
\[ \pi^*{(s)} = \argmax_{a \in A} V^*(s). \]

There are two main approaches to solving MDPs, depending on whether or not the reward function and transition function of the MDP are known. In problems in which both functions are available, an agent can use \emph{planning} methods to directly calculate an effective policy by computing the optimal value of each state and then the optimal action~\cite{bellman1966dynamic}. More specifically, these methods typically involve calculating the optimal value function and then the optimal policy by using dynamic programming or linear programming. However, in problems in which either or both of these functions are unavailable, an agent can use \emph{reinforcement learning} methods to gradually learn an effective policy by performing actions and observing rewards to estimate the optimal value of each state and then the optimal action~\cite{sutton1995learning}. That is, all reinforcement learning is built upon MDPs and their variants. In particular, these methods usually involve estimating the optimal value function and then the optimal policy by interleaving greedy actions with exploratory actions.\footnote{Although we do not discuss many solution methods in this work, many, such as value iteration~\cite{bellman1966dynamic}, RTDP~\cite{barto1995learning}, Monte Carlo tree search~\cite{browne2012survey}, Q-learning~\cite{watkins1992q}, and SARSA~\cite{chen2008least} have proven to be effective across a variety of applications, including Atari~\cite{mnih2015human}, chess~\cite{silver2018general}, and StarCraft~\cite{vinyals2019grandmaster}.}

{\bf Example: } Consider a power plant that supplies power to several neighborhoods. The goal of the power plant is to balance three potentially competing objectives: it must (1) supply power to each neighborhood as cheaply as possible, (2) avoid outages, and (3) reduce excess power that is stored in its battery network and dissipates gradually. Armed with the MDP framework, we can formally represent the decision-making problem of the power plant as an MDP $\langle S, A, T, R \rangle$. In particular, suppose the plant can supply a maximum of $R$ kilowatts (kW) to a set of neighborhoods $N$ where each neighborhood $N_i$ demands $D_i$ kW. The plant incurs a cost of $C \geq 0$ per kW generated and charges each neighborhood a price $P \geq C$ per kW. It also incurs a cost $E \propto R-D$ for generating excess power. We assume the power plant either meets all or none of the power demand $D_i$ kW for a given neighborhood $N_i$: that is, the plant supplies either $D_i$ or $0$ kW to neighborhood $N_i$. Thus, our set of states $S = E \times P \times D_1 \times \dots \times D_{|N|} \times F_1 \dots \times F_{|N|}$ where $P = \{ \textsc{Low}, \textsc{Normal}, \textsc{High} \}$ is the current price of power, $D_i$ is the current power demand for the neighborhood $N_i$, and $F_i = \{ \textsc{Fulfilled}, \textsc{Unfulfilled} \}$ is the current fulfillment status of the neighborhood $N_i$, reflecting whether or not the current power demand $D_i$ is met.

The plant has two ways to control load: it can increase or decrease the price $P$ in order to keep total demand $D = D_1 + D_2 + \dots + D_{|N|}$ close to, but below, the maximum rate $R$. However, if $D > R$, the power plant can also terminate supply to neighborhoods $\bar{N} \subset N$. The set of actions is thus $A = \{ \oplus, \ominus \} \times \mathcal{P}(N)$, where $\oplus$ and $\ominus$ increase and decrease the current price of power $P$ and the powerset $\mathcal{P}(N)$ is every combination of neighborhoods for which power can be shut down.\looseness-1

The transition function $T(s, a, s')$ represents how the probability of the power demand of each neighborhood varies with the current price of power. The reward function $R(s, a)$ represents the relative cost of service interruptions, charging a price of power higher than the cost of power generation, and having to store excess power in a battery network. A non-myopic model like an MDP is obviously preferable to a classifier in this decision-making scenario since the outcome of a given action has both some uncertainty as well as some impact on possible subsequent actions.

{\bf Frontiers of Sequential Decision Making: } A substantial body of work focuses on solving MDPs efficiently given that the computational complexity of solving them ``blows up'' with the size of their states and actions. This problem is colloquially referred to as the \emph{curse of dimensionality} in AI literature. To provide some background, we highlight three common approaches to solving MDPs approximately. \emph{Approximate programs} estimate the optimal value function and then calculate the optimal policy for that estimated optimal value function by using approximate forms of dynamic programming~\cite{bertsekas2011approximate,powell2016perspectives} or  linear programming~\cite{guestrin2003efficient,petrik2009constraint,poupart2015approximate,malek2014linear}. \emph{Replanning methods} generate a policy for a subset of states (called a partial policy) and then generate a new partial policy whenever a state is encountered for which the partial policy is undefined~\cite{smith2006focused,yoon2007ff,pineda2017fast}, enabling the solver to reason only about the most likely states. Finally, \emph{abstraction methods} build an abstracted MDP to reduce the size of its state and action spaces and then solve for the optimal policy of the abstracted MDP~\cite{dean1997model,ravindran2002model,givan2003equivalence,ferns2004metrics,li2006towards,yu2009basis,biza2018online,saisubramanian2019adaptive,nashed2021solving,nashed2022selecting}, retaining relevant details and condensing those less important. In practice, approximate MDP solvers may employ various combinations of these approaches.

In addition to work on solving MDPs efficiently, there are many MDP extensions that can represent different classes of decision-making problems. Here we focus on MDPs, a model for decision-making problems in which the current state can be directly observed. However, there are many decision-making models with different forms of expressiveness for decision-making problems with different properties. For example, for problems in which the current state is not directly observed by the agent, requiring the agent to manage a belief over the current state, we can use a partially observable MDP (POMDP)~\cite{kaelbling1998planning}. For problems in which the agent must find the shortest path from a start state to a goal state, we can use a stochastic shortest path problem (SSP)~\cite{kolobov2012theory}. For problems in which multiple, decentralized agents must coordinate, we can use a decentralized MDP (DecMDP)~\cite{bernstein2002complexity}. There are many other MDP flavors, but they also suffer from the curse of dimensionality, often so much so that they require specialized approaches to solve efficiently.

{\bf Fairness in Sequential Decision-Making Systems: } Recently, there have been arguments for using MDPs to model decisions traditionally handled by supervised learning~\cite{heidari2021allocating,zhang2021recommendation}. However, there are relatively few efforts at producing fairness definitions consistent with the definition of an MDP, such as  \citet{wen2021algorithms}, who use constrained MDPs to express fairness constraints for a subclass of MDPs with separable reward and transition functions. Here, expected reward (value) plays an analogous roll to the loss function in supervised learning. Some surveys also highlight the temporal nature of the many decisions AI systems make, but focus primarily on allocative tasks, stopping short of expanding these problems to encompass the types of sequential decision-making models most often deployed by embodied AI systems \cite{zhang2021fairness}.\looseness-1

One class of MDPs that is relatively well-studied, however, is the multi-armed bandit problem~\cite{bouneffouf2020survey}. In this problem there is a set of arms (actions), each yielding a different reward according to an unknown distribution. The objective is to determine the arm to pull that maximizes expected cumulative reward. Formally, the multi-armed bandit problem is a class of MDPs in which the agent performs a single action instead of a sequence of actions. Here, recent work has offered models and algorithms for introducing different notions of fairness~\cite{metevier2019offline}. \citet{joseph2016fairness,joseph2016fair,joseph2018meritocratic}~initiated this line of research by introducing a meritocratic definition of fairness that ensures that a better arm is always favored over a worse arm despite uncertainty over each arm's expected reward. Then, extending this work to each arm's reward distribution instead of its expected reward, \citet{liu2017calibrated}~offered a method for ensuring that two arms are pulled roughly the same number of times if they satisfy a notion of similarity based on these reward distributions. Moreover, in the context of fairness constraints, there has been a range of methods for ensuring that each arm is selected a minimum number of times~\cite{chen2020fair1,chen2020fair2,claure2020multi,patil2021achieving,li2019combinatorial}. Finally, as a way to reason about group fairness, \citet{schumann2019group}~offered a method for partitioning the arms into different sensitive groups based on protected features, such as race, age, and socio-economic status, that are in turn picked from according to a given definition of fairness. However, while these works examine fairness in the context of multi-armed bandits and have led to encouraging results, it remains challenging to extend these ideas to MDPs because MDPs are a strict generalization of multi-armed bandits in which the agent must optimize over a sequence of actions instead of a single action, and each action affects state transitions.

\section{Conceptualizations} \label{sec:conceptualizations}

In this section, we briefly examine how the fairness and ethical decision-making literatures have conceptualized their work. Fairness is an essentially contested construct \cite{hutchinson201950,jacobs2021measurement}, and fairness in predictive systems, though generally centered on unequal exposure of certain people to potential system failures, has been conceptualized in a variety of ways, among them individual and group fairness. Individual fairness requires that similar individuals be treated similarly \cite{dwork2012fairness}, whereas group fairness requires that different groups be treated similarly. As \citet{jacobs2021measurement} explain, debates about individual versus group fairness, as well as debates about the right definitions of individual and group fairness, reflect an array of ``different theoretical understandings'' of what constitutes fairness. For example, some definitions of group fairness reflect concerns that predicted scores should have the same meaning for people from different demographic groups, while others reflect concerns that errors experienced by people of different groups should be comparable \cite{jacobs2021measurement}. \looseness=-1

Despite these differences and the substantial debate they have engendered, conceptualizations of algorithmic fairness reflect some general patterns. For example, most fairness operationalizations formally represent only decision subjects; other stakeholders impacted by system predictions, such as business owners trying to make hiring decisions or the dependents of someone eligible for parole, are rarely represented. Much of the fairness literature has focused on settings where systems might withhold resources or opportunities from some people, such as recidivism \cite{angwin2016machine}, hiring \cite{sanchez2020does,liu2020disparate}, and education~\cite{heidari2021allocating,mouzannar2019fair,marcinkowski2020implications}, rather than settings where systems might represent some people unfavorably.\footnote{\citet{barocas2017problem} and \citet{crawford2017trouble} refer to these as allocative and representational harms, respectively.}\footnote{This is not to say, of course, that there has not been ample work examining the latter, for example \citet{sweeney2013discrimination} on discrimination in online ads and \citet{noble2018algorithms} on search engines' reproduction of racial and gender stereotypes.} These focuses reflect, as \citet{hoffmann2019fairness} writes, ``liberal anti-discrimination discourses in the law, which have historically sought to address injustices in the distribution and exercise of important rights, opportunities, and resources in domains like voting, housing, and employment.'' In addition to anti-discrimination law, conceptualizations of fairness often draw on political philosophy, particularly theories of distributive justice \cite{binns2018fairness,heidari2019moral,kasirzadeh2022algorithmic}. Critiques of the fairness literature have observed that it may leave assumptions about what constitutes fairness unexamined \cite{jacobs2021measurement}, treat fairness as a self-evidently appropriate framing (as opposed to other possible framings, such as justice) \cite{bennett2020point}, or place too much faith in a fairness framing's ability to address longstanding structural concerns \cite{hoffmann2019fairness}. Work addressing the ethical implications of predictive systems from perspectives other than fairness has also emerged, including analyses of systems' underlying logics and their sociohistorical contexts \cite{stark2022physiognomic}, how systems reproduce patterns of injustice or power relations \cite{keyes2018misgendering,mohamed2020decolonial}, the values and incentives of the disciplines producing systems (e.g., machine learning) \cite{birhane2022values}, and the labor upon which systems rely~\cite{irani2013turkopticon}.\looseness-1

By contrast, within the ethical decision-making literature, concerns about system outputs manifest from two distinct scenarios. First, how might a sequential decision-making system cause harm due to an inadequate decision-making model? For example, if the decision-making model from $\S 2$ had only two options for price, \textsc{High} and \textsc{Low}, customers may be charged more than necessary in scenarios where the optimal action is to set the price to \textsc{Normal} instead of \textsc{High}, since the decision-making model does not recognize this as a possibility. Similarly, if the number of neighborhoods used to model an area decreases, then some people may be left without power even when not strictly necessary since the agent cannot make more fine-grained decisions. Note that the fix for both of these examples is to make a more complex, and therefore more expensive, decision-making model. Second, how should the system behave when faced with a decision for which there is no obviously good outcome, or a high degree of risk? For example, if the power management agent must cut power to a neighborhood, how should it decide which neighborhood to cut? Some of the first ethical decision-making proof-of-concept systems were focused on military applications where there was potential for lethal use of force \cite{arkin2008governing,arkin2009ethical,arkin2009governing}. This application provides a context of rules for the moral conduct of warfare, studied extensively by ethicists and moral philosophers, which influence the way many scholars view ethical decision making today.

Ethical decision-making systems are considered ethical when their behavior aligns with a set of rules for acting, either learned or prescribed. These sets of rules are devised under the assumption that systems which follow those rules will minimize harm. This conceptualization sits in stark contrast to that of the fairness literature, where many fairness definitions are expressed in terms of relative failure rates or unevenly distributed system error.\footnote{\citet{raji2022fallacy} observe that fairness research may often neglect to ask whether systems function in the first place.} In some sense, strict adherence to a set of rules for acting sets a much higher standard for agent behavior, though in practice expressive, effective, and general rule sets are exceedingly difficult to generate. Justification for this strategy often comes from moral philosophy, where ethical theories are broadly understood to provide such rules for acting. Several major ethical theories have been used to motivate autonomous systems where an agent is, morally speaking, required, permitted, or prohibited from taking specific actions in specific states depending on whether that action in that scenario violates the rules of the ethical theory. These theories include Act Utilitarianism~\cite{anderson2005towards,grau2006there,kim2018computational}, Kantianism~\cite{powers2006prospects,wiegel2009combining,hooker2018toward}, Virtue Ethics \cite{kulicki2019virtue,neubert2020virtue,svegliato2021ethically}, Norm-based systems \cite{kasenberg2018norm,frazier2019learning}, The Veil of Ignorance \cite{leben2017rawlsian,nashed2021ethically}, Divine Command Theory \cite{bringsjord2012divine}, The Golden Rule \cite{nashed2021ethically}, and Prima Facie Duties \cite{anderson2011prima,svegliato2021ethically} among others. In addition to these applied works, there have been many more theoretical pieces examining when and why particular ethical frameworks ought to be used~\cite{goodall2014machine,powers2005deontological,powers2006prospects,hibbard2012avoiding,farina2022ai,peeters2021designing,liu2017confucian,zhu2019confucian,vandemeulebroucke2018use,yew2021trust}. However, these systems are still largely imagined, and we are not aware of any real-world systems yet in operation, in contrast to ``fair'' predictive systems which we know operate in a variety of public settings already.\looseness-1

\textbf{3.1) Conceptualizations of the ``right thing'' differ fundamentally between fairness and ethical decision making.} Fairness research often conceptualizes doing the right thing as equalizing exposure to various failure modes or performance metrics in aggregate across subsets of the population delineated by protected attributes. By contrast, ethical decision-making research typically conceptualizes doing the right thing in terms of adherence to a set of rules applied to all possible scenarios, which often follow a logical framework from moral philosophy.

\textbf{3.2) Conceptualizations of ethical behavior for sequential dec-ision-making systems are shaped by models' increased capacity for reasoning.} Unlike predictive models, sequential decision-making models allow a system to reason explicitly about the effects of its actions, including long-term consequences. Thus, sequential decision-making systems are generally conceptualized as systems that act---and enact change---in the world, shaping what it means for a these systems to behave ethically. This represents a fundamentally different perspective on a system's role within its sociotechnical context compared to predictive systems \cite{alaieri2016ethical,yazdanpanah2022reasoning}. For example, many discussions of ethical decision making focus on long-term behavior \cite{nashed2021ethically,svegliato2021ethically}, whereas many conceptualizations of fairness incorporate no notion of either a system's future decisions or the downstream consequences of those decisions. Recently, \emph{welfare}, which generally measures holistic outcome effects, rather than error rates, has been proposed as an alternative measure \cite{hu2020fair,finocchiaro2021bridging,kallus2021fairness}, bringing evaluation ethea in ethical decision making and fairness slightly closer. Specifically, framings using welfare allow detailed longitudinal analyses previously scarce in fairness literature, and suggest conceptualizations of predictive systems in their contexts of use rather than as standalone systems.

\textbf{3.3) While concerns in ethical decision making are rarely articulated in terms of fairness, fairness may nevertheless offer a useful lens for evaluating the outcomes of deployed sequential decision-making systems.} For example, the MDP given in \S2 may be more likely to interrupt service to neighborhoods with a particular demographic. We will discuss potential underlying causes and intervention strategies in later sections, but here we simply highlight that at a high level, fairness-type audits of such a system could potentially detect this type of behavior and that research on how to do these types of audits for sequential decision-making systems is absent from the literature. Similarly, research addressing \emph{if and when} fairness is the right construct for analyzing sequential decision-making system outcomes is also absent.\looseness=-1

\section{System Development} \label{sec:pipelines}

In this section, we examine common prediction and sequential decision-making pipelines, cover some important differences---including system inputs and outputs, how expert knowledge is encoded, and informatic assumptions---and discuss what these differences suggest as objects of analysis for future research. Predictive systems have two components: data and a function approximator. The goal is to learn a function that can predict some hidden variable using data. Ideally, the data is accurately labeled, representative of the deployment setting, and plentiful enough to train a model. These assumptions of course may not all be met by development or deployment conditions, and while these are common topics of research in the broader machine learning community, fairness researchers have also proposed methods for handling flawed data \cite{buolamwini2018gender,chen2019fairness,wang2021fair,madras2019fairness,awasthi2021evaluating}, out of training distribution data \cite{taskesen2021statistical,singh2021fairness}, and efficient learning \cite{slack2020fairness}. 

\textbf{4.1) By contrast, the product of a sequential decision-making system is a policy that when executed results in a sequence of actions taken in the world.} Instead of function approximators, sequential decision-making systems use planners. Often, these planners produce \emph{provably optimal} policies, meaning that, w.r.t. its model, the agent maximizes its cumulative expected reward. The existence of a policy instead of a set of i.i.d. decisions means that understanding system behavior is more contextual and not always possible with the same statistical measures. Because policies represent situation-dependent prescriptions for actions, analyzing a policy requires inspecting the action that would be prescribed by the policy for every state. This represents a major departure from aggregate measures of monitoring behavior.

\textbf{4.2) When designing a decision-making model, developers hypothesize about the structure and value of unseen data, rather than extracting patterns from existing data.} In other words, developers write down what they think is (or will be) true about the world~\cite{schaefer2005modeling,boucherie2017markov}. For example, when specifying the reward function,  they decide the relative utility of outcomes; when enumerating the state space, they forecast the importance and availability of different data. Therefore, the most important and most fallible assumption in sequential decision making is that the model is faithful enough to the dynamics of the real world to support effective decision making. For example, if the transition function representing changes in demand is not perfect, then there may be scenarios when the agent's actions are optimal w.r.t. the model, but not the real world.\footnote{It is possible to learn or refine the transition and reward functions from data. Reinforcement learning is used to learn the transition function by sampling actions and inverse reinforcement learning is used to learn the reward function by observing sequences of state-action pairs from an agent running a policy.} Problems may also arise from under-specified state spaces, as in the power grid example from \S3. Adding more state factors or increasing their domains creates exponentially and polynomially more states, respectively. This tradeoff between model descriptiveness and computational tractability is not present in predictive systems since typically defining the desired classes or the meaning of the numerical output is not a fundamentally intractable problem.\footnote{While increasing the number of classes does increase the data required to learn robust models, a larger space of class labels alone does not make the problem computationally intractable.} 

While tricky to get right, the practice of crafting models has several advantages compared to learning from data alone. The ability to provide an initial model of the world, even if refined using data, is useful as it allows developers to encode knowledge and model feedback effects that otherwise may be difficult to learn. Thus, developers spend considerable effort in creating decision-making models that are as compact and descriptive as possible, and experts are highly valued for their ability to design tractable, accurate models. Often, these models also require specific domain expertise, and while there are some cases in which data may be gathered to enhance model building or, most often, to improve the transition function, generally data is not available for the task at hand, and in some cases relevant data may be unavailable altogether.

\textbf{4.3) The designs of states, actions, and rewards are obvious objects of analysis for decision-making models.} The process of defining the state and action spaces and reward functions can be thought of as a structured way of encoding expert domain knowledge. Through implicit assumptions about how the world works, the purpose of the system, the source of data, or the responsibility of the agent for certain outcomes, developers encode expert knowledge about decision-making scenarios. This knowledge is not only important, but required, in order to make problems manageable computationally. However, it is also through these mechanisms that decisions are made which may cause harm. 

For predictive systems, expert knowledge is often exploited via careful curation and selection of data, or through choosing the spaces of inputs and outputs (labels, classes, or target variables). The space of outputs is typically driven by the task. It may map directly from a description of the decision to be made, such as whether a manufactured part passes a quality control test or not, or it may correspond to a component of a larger decision-making task, such as calculating recidivism risk which aids in determining bail. The set of inputs is often more difficult to determine and has historically been more controversial in the fairness literature \cite{khani2021removing} than in the ethical decision-making literature. First, we should note that all features are proxy data for the variable of interest. If we could measure the variable of interest directly, there would be no need for a prediction system. So, all inputs to such prediction models are pieces of data that modelers believe either cause, or at the very least correlate with, the variable of interest. The existence of this choice creates two concerns. First, from an engineering perspective, the choice of inputs can be somewhat of a ``dark art'', and even with the advent of deep learning it is still often challenging to understand if a function is challenging to learn because of uninformative features or some other phenomenon. Second, from an ethical perspective, including data that is unlikely to be causal and may be correlated with data protected under non-discrimination laws, can lead to concerns that labels are being assigned in part due to race, gender, or some other protected attribute. This is especially difficult when there are proxy variables that correlate with both the variable of interest and protected variables.\looseness=-1

Sequential decision-making systems offer analogous choices, corresponding to the choice of state and action spaces and the reward function. The action space, like the space of labels, is typically more straightforward to conceive and less controversial, since it simply represents the capabilities of the agent. For example, there are a limited number of commands an autonomous car can give its motors and thus it is usually clear what is appropriate and what is not.\footnote{This is less clear in systems that use hierarchies of sequential decision-making systems \cite{guestrin2012distributed}, but the vast majority of real-world applications do not abstract sequential decision making to more than one level.} The choice of state factors on the other hand creates two concerns. First, from an engineering perspective, the tradeoff between computational tractability and expressive decision-making power is hard to get right. While the complexity of solving an MDP is polynomial in the size of the state space, the size of the state space scales exponentially with respect to the number of state factors. This often severely limits the number of state factors that an agent can use for decision making. The ability to create state spaces that are small enough to solve policies for, but also do not obfuscate important nuance by abstracting away important details of the situation, is also something of a ``dark art''.

From an ethical perspective, both the state space as well as the reward function may be poorly designed and cause harm. For example, given a transition function that describes the \emph{true} probability of a power shortage in a certain neighborhood, the MDP may make a decision to cut power or raise rates in order to balance the load of the entire network. This true probability may be accurate, and the MDP may be taking the optimal action with respect to both its model and the real-world application, but this probability may also be influenced by societal forces, like neglect of local infrastructure, that correlate with a sensitive attribute, such as race. Behaving in accordance with certain ethical norms may create need for additional state factors beyond those required to complete the task. If this additional detail is not encoded in the state space, the resultant policy will not be able to distinguish between scenarios where multiple actions are roughly equivalent with respect to the task but have drastically different ethical implications. Even given a descriptive and compact state space, designing a decision-making agent that behaves ethically requires a reward function. The reward function presents another unique challenge in that it implicitly combines and homogenizes all possible outcomes onto the same numerical scale. Thus in a regular MDP, no matter how large and descriptive the state space, all good and bad outcomes, regardless of how they might be measured and who or what they might affect, are converted into the same ``unit''. There are many philosophical debates about whether this is even possible to do in a principled manner~\cite{kolm1993impossibility,shafer2009fundamentals,wedgwood2022reasons}, but setting aside that controversy, this still challenges developers and creates a potential source of error. Finally, although the true transition function is fixed given the state and action spaces, and thus not usually considered a design choice, most MDPs do not have perfectly accurate transition functions and inaccurate transition estimates can also lead to undesired behavior.

\textbf{4.4) Design processes for decision-making models often lack scrutiny.} As with prediction, the ``dark art'' of the design process, particularly regarding choice of state factors, means it is generally not a topic of discussion in the research community, let alone available for public scrutiny. Much of the design process is done by instinct, relying on domain experts, and is not well-codified or written down. Reasons behind decisions are often not available via publications or other documents, and the final models, if believed to be interpretable in their own right, may be taken as self-evidently appropriate and therefore under-inspected. This pattern is exemplified here \cite{keizer2013training,de2020real,de2020ethical}, where the relevant variables are simply stated without further explanation or justification. Because of the proof-of-concept nature of most existing work on ethical decision making, the pattern of implicit justification extends to this research as well. Most works do not use metrics based on specific attributes and instead examine how to specify high-level abstract rules. In these cases, the justification for these omissions is that the particulars of the state factors or rewards are simply placeholders used to study the effect of the rule, for example~\cite{winfield2014towards,vanderelst2018architecture,dennis2016formal,van2003contextual}. However, as researchers begin designing systems for more specific applications and with intention to deploy them, this justification will need to be critically examined.\looseness=-1

Given these observations, a number of research questions immediately suggest themselves. For example, for a particular decision-making model, which stakeholders are explicitly modeled? What kinds of approximations of the world are common, and what assumptions underpin them? Whose domain expertise is solicited? Are model aspects borrowed from one application or deployment setting to another, or always developed afresh? How does model design take into account the larger systems that models participate in? We are not aware of research that explicitly studies processes related to the design and development of \emph{sequential decision-making systems}, although there are substantial bodies of adjacent literature on interpretability \cite{petrik2016interpretable,brown2018interpretable,topin2021iterative,miura2020maximizing}, explainability~\cite{bertram2018explainable,juozapaitis2019explainable,nashed2023causal}, and generally on participatory design~\cite{martin2020participatory,liao2019enabling,madaio2020co}.

\section{System Evaluation and Measurement} \label{sec:measurement}

Ideals regarding how a system ought to behave in the abstract are only as good as our ability to show, either theoretically or empirically, that they will adhere to these ideals when deployed. This is a challenging task for several reasons. To begin, what is right or wrong, or what is desired behavior, is not always agreed upon by all stakeholders. This has been discussed at length in the fairness literature \cite{corbett2018measure,ekstrand2018all,kasy2021fairness,friedler2019comparative,yeom2021avoiding,harrison2020empirical,hutchinson201950,cohen2022price,mashiat2022trade,zhang2022affirmative,baumann2022enforcing,pfohl2022net,lum2022biasing} and in practice leads to uncertainty regarding measurement and evaluation. Research on operationalizing and comparing different definitions and measures of fair or ethical behavior makes up a significant percentage of papers published at FAccT and similar conferences. Here, we will explain and compare the different metrics, tools, and strategies employed by researchers in fairness and ethical decision making as they attempt to show that their system will ``do the right thing''.

For predictive systems, measuring fairness means operationalizing some conceptualization of fairness, typically by statistically analyzing a system's performance over one or more groups of users \cite{hutchinson201950}. These measurements might look for predictive parity \cite{celis2019classification,dieterich2016compas,agarwal2018reductions}, error rate balance \cite{iosifidis2019adafair,zhao2019conditional}, or anti-classification \cite{zafar2017fairness,jiang2020wasserstein,kamishima2018recommendation}, and are often focused on counterfactual analysis, either at the group \cite{dwork2018decoupled,speicher2018potential,kusner2017counterfactual} or individual level \cite{jiang2020wasserstein,narasimhan2020pairwise}. Evaluating predictive systems with these measures requires access to the predictive model, and either real-world data or high-quality simulated data. When data is readily available, only a computer capable of running the model is required for evaluation. Whether these measures ultimately track the quality of the outcomes for users, however, is still an open question \cite{glymour2019measuring}.

\textbf{5.1) Sequential decision-making systems must be deployed to fully evaluate them.} While predictive systems are similar in that some harms may only be evident in deployment, when the system is viewed holistically in its social context and the impact of decisions is clarified, researchers can at least take measures of fairness absent a deployed system. However, shortcomings of decision-making models are nearly impossible to uncover without an agent operating in the real world, taking actions, affecting its environment, and encountering real-world data from the resultant states. Thus, the only way to systematically understand harmful outcomes is to deploy, which is expensive, time-consuming, and often unsafe. To emphasize, most policies are optimal w.r.t. their model and thus abide by constraints of their model. However, assumptions made by the model may not reflect the real world and thus lead to unintended behavior.

\textbf{5.2) Sequential decision-making models are often embedded in larger systems.} Often, MDPs are part of a larger system, such as a robot~\cite{brunke2022safe,lauri2022partially,ali2020path,suarez2019practical}, and it may be challenging to write a reward function that represents its high-level task, which may be a mixture of several objectives. Thus, we often evaluate these decision-making models using a task-based metric~\cite{canal2019probabilistic,lacerda2019probabilistic,ali2020path,smith2021socially,schillinger2021adaptive}. For example, consider a robot running a policy for loading boxes into a truck. We can compare policies generated by different decision-making models, regardless of how their reward functions represent the task, simply by counting the number of boxes loaded into the truck. This seems simple, but is often the most costly and tedious experiment to run since it requires a fully functioning system, whereas theoretical analysis and simulation do not. Moreover, this makes predicting the ethical impact of different interventions even more challenging, and compounds the problem of not being able to test without fully deploying. Given the high costs and risks associated with the process, how might we safely approach system evaluation for ethical concerns, and how might we incentivize doing so?\looseness-1

\textbf{5.3) Though imperfect, many proxy measures for policy quality exist.} Exact planners are optimal so we rarely evaluate the planning algorithm, but rather the decision-making model and its ability to produce accurate real-world decisions. One common technique is to simply spot-check the policy at different states where the agent is balancing competing reward signals to verify it behaves as expected. A more methodical strategy is to calculate the probability of reaching a certain bad state if the agent follows the optimal policy and begins in a given state. For example, we could compute the probability that neighborhood $N_i$ experiences a service interruption during the next year. This type of analysis can uncover some policy errors, but is limited due to (1) the difficulty in enumerating all bad states or outcomes and (2) humans' poor intuition for likelihoods of different events. In short, sanity checking policies in this way is time-consuming and prone to error.\looseness-1

Moving towards in situ evaluation, there are several methods for evaluating policies that rely on the ability to simulate deployments. One basic method is to simulate the agent executing a policy many times and calculate the variance in performance in an effort to understand its reliability. A more principled method, with a prediction analogue, is to test the performance of the policy under a variety of transition functions, or ``possible worlds", typically in terms of the total reward earned. For example, we may be uncertain about whether our transition function correctly represents the probability of change in demand following a change price action. That is, we are not questioning whether the outcome is stochastic, we are already modeling that; we are questioning whether our model correctly captures the relative probabilities of different events, and the robustness of our policy to potential errors in the model. This is similar to some prediction or transfer learning settings where training, testing, and validation data sets represent different distributions of data~\cite{nanda2021fairness}. MDPs which are solved assuming a distribution over transition functions are called robust MDPs \cite{ben1998robust,nilim2005robust}. There is also a large body of related work on ``safe'' policies or ``safe'' learning, where ``safety'' has been defined in terms of behavioral constraints~\cite{wachi2020safe,saisubramanian2021multi}, policy ergodicity \cite{moldovan2012safe}, risk metrics \cite{ruszczynski2010risk,feyzabadi2014risk}, and the probability of improving a policy \cite{thomas2015high}. 

Finally,\footnote{We should note for completeness that there are many approximate techniques for solving sequential decision-making problems. When evaluating these techniques, for a fixed decision-making model, directly measuring the value function can often provide a signal as to the quality of the resultant policy since all value functions are upper bounded by the value function induced by the optimal policy---the policy we would get if we used an exact planner. Even better is measuring the actual cumulative reward experienced by the agent using simulation or data collected from a deployment. However, if the model is changed in any way, including the discount factor, these comparisons cannot be run across models. This is because changes to rewards, transition probabilities, or the discount factor can change the scale of the value function, producing different upper bounds and therefore preventing fair comparison.} one important difference between prediction problems and sequential decision problems is that when deployed, predictive models generally cannot know whether the result of their inference is correct. In contrast, sequential decision-making systems always know that the chosen action was optimal in expectation w.r.t. the model. Moreover, they can immediately observe the reward for a given outcome, even if it is unexpected. What is unknown is whether the optimal decision w.r.t. the model is also the optimal decision for those affected by the decision. Of course, the reward may not be perfectly aligned with preventing harms, but the ability to examine performance longitudinally is a powerful benefit.\looseness-1

\textbf{5.4) Methodical evaluation of sequential decision-making systems for ethical behavior is an open problem.} We are not aware of rigorous empirical research on harms produced by deployed sequential decision-making systems, including basic questions such as ``Who is harmed?'' The question is more often framed in terms of rules violations, but even these studies are not common due to the more theoretical nature of most existing research and lack of access to sequential decision-making systems. Although there are many methods for evaluating policies w.r.t. their decision-making models, ethical decision-making researchers are almost completely in the dark when it comes to understanding the impact of their agents on the world outside the model. We view this as a critical shortcoming in existing research.\looseness-1 

\textbf{5.5) Currently, auditing sequential decision-making systems poses a serious logistical challenge to researchers.} In order to audit most sequential decision-making systems, an auditor would require the physical agent, the agent's policy, and any supporting software that connects the two, such as algorithms for determining states from data and controllers for executing actions specified by the policy. For example, auditing decision making in an autonomous car would require the car and its entire software stack in order to verify how it behaves in different scenarios. This is a significant obstacle for researchers interested in transparency and accountability. Moreover, since these systems are often functioning as part of a larger system, it can be difficult for the public or regulators to even know where systems are deployed. Academic~\cite{wilson2021building} or community audits, such as audits of commercial image cropping algorithms \cite{birhane2022auditing,chowdhury2021introducing}, are challenging if not impossible.

\section{Mitigations} \label{sec:solutions}

Often, the goal is to modify, augment, or in some other way intervene in the decision-making process of an existing system in order to ensure that it behaves fairly or ethically. In this section, we examine common interventions, including data augmentation, reward modification, optimization formulation, and system integration. For predictive models one of the simplest interventions is to collect or generate more data (data augmentation), under the assumption that as more samples are acquired the training set will improve its approximation of the true distribution and the model will learn a more accurate, representative function \cite{sharma2020data,zhang2020towards,vannur2021data,pastaltzidis2022data,rateike2022don,wu2022fairness}. This practice works well if data deficiency is the \emph{only} reason for poor performance. In many cases, the problem is not the amount of data but the quality of the data. Specifically, there are often artifacts in the data, such as correlations between attributes like race or gender and the target variable, that we do not want our predictive model to learn. One way to mitigate this is to try to balance the data set to remove these correlations within the data by adding new data points or editing existing ones (data curation) \cite{cao2022fair,leavy2021ethical,cai2022adaptive}.

Data-based interventions present challenging tradeoffs. For example, gathering sensitive data may be required to ensure a balanced data set with respect to certain attributes, or verify a model meets certain fairness criteria. However, doing so raises privacy concerns \cite{pujol2020fair,ekstrand2018privacy,cheng2021can} as well as accuracy concerns when sensitive attributes are not readily available or easily identifiable \cite{ghazimatin2022measuring}. Moreover, many researchers have rightly pointed out that common operationalizations of race, gender, and other socially constructed concepts may in fact be more harmful than helpful \cite{green2020algorithmic,kasirzadeh2021use}. Nonetheless, this is often the only data available to these systems. There is somewhat of a paradox in wanting to avoid sensitive attributes influencing predictions and reifying certain categories, but requiring these attributes in order to verify these criteria \cite{andrus2022demographic}.

\textbf{6.1) Data augmentation and curation are often impractical or unsafe for sequential decision-making systems.} While it is possible to use data from actual deployments along with reinforcement learning to improve sequential decision-making systems, this is often very costly and occasionally unsafe. Generally speaking, MDP agents cannot be developed in isolation with data from elsewhere. The agent itself is required in order to generate data by interacting with the world---taking actions, recording state, and experiencing reward. Many researchers get around this problem using simulation, but again, many types of failures may occur in the real-world that do not show up in simulation, especially those related to ethical behavior. Thus, real-world deployments are often a bottleneck for gold-standard evaluation.

\textbf{6.2) Computation constraints limit performance of sequential decision-making systems.} While there is no direct analog of data augmentation or curation in MDPs, the decision-making models themselves are often augmented by expanding the state space. This is done by either adding new state factors or expanding the domains of existing ones, for example by adding new neighborhood factors which represent subsets of the original neighborhoods in the power management problem. This delineates some scenarios that were previously considered identical, allowing the agent to choose different actions under those conditions. By adding new neighborhoods, the agent has more fine-grained control over service interruptions and can maintain power to more homes. We may also add completely new state factors to the MDP, such as the existence of backup generators in some neighborhoods, that can be used to modify the reward function so that it reduces penalties for outages in these neighborhoods since the ultimate impact is reduced. Thus, larger, more descriptive state spaces often produce more nuanced, performant policies at the cost of time required to generate a policy.

Generally, in prediction, more computation cannot improve the estimate of a target variable. This is not so for sequential decision making. While MDP\footnote{We should emphasize the tremendous volume of work on extending MDPs to other informatic settings. For example, state factors may not be directly observable \cite{kaelbling1998planning}, such as when a pedestrian becomes temporarily occluded from the view of an autonomous vehicle. Their position is unobservable, so the vehicle maintains a belief over the pedestrian's location and thus a belief over the state of the world. Other specialized models have been developed for decentralized behavior \cite{bernstein2002complexity}, adversarial scenarios \cite{gallego2019reinforcement}, and hierarchical decision processes \cite{hauskrecht2013hierarchical} among many, many others. These models vary greatly in their assumptions and complexity, and understanding the feasibility of different interventions across different models is an open question.} solvers have polynomial complexity in the number of states, the number of states often grows exponentially with the number state factors. This presents a challenge as improvements via state space augmentation are limited since adding state factors orthogonal to the task adds exponential cost. Moreover, additional states which do not map to different actions increase computational cost without increasing performance. In practice, decision-making models are often \emph{necessarily} approximations, which marginalize (in the computational sense) some variables or compress different scenarios into the same state representation to reduce model size and therefore compute load. 

\textbf{6.3) There are no existing methods to account for how decisions regarding the state space or reward function affect stakeholders in the general case.} State factors are not restricted to the data at hand---they may represent any information the agent can measure or sense. In some cases, this prevents models from reproducing historically biased patterns since unwelcome correlations simply are not present. Furthermore, well-designed MDPs only use state factors that are important to making decisions for the task. Typically, race, gender, and other sensitive attributes not related to the decision may be omitted. While this does not guarantee ethical behavior, these traits reduce the severity of some problems common with predictive systems. An important corollary to this is that fairness is usually not useful as an optimization constraint since protected attributes are typically not encoded in the state space. 

However, there may still be insidious correlations, and identifying them is one area where the ethical decision-making community could benefit from the experience and insights from the fairness community. To date, there is little research on whether decision-making models exhibit the same problems as predictive models, such as when removing protected attributes explicitly from the reasoning process does not prevent differential treatment with respect to those attributes. Moreover, there are many broader questions remaining about what it means for protected attributes to be part of the reasoning process in sequential decision making. For example, even if protected attributes are not represented as state factors, can the transition and reward functions still encode harmful patterns? How might state factors implicitly encode sensitive attributes? How might sequential decision-making outcomes reproduce patterns of discrimination? Simply adding protected attributes as state factors seems ill-advised since unless these factors affect the reward or transition functions they will not affect the reasoning process. Understanding how changes to a decision-making model may affect resultant policies is an important part of designing MDPs. However, even for specific applications there is essentially no systematized or codified method for predicting the impact of model changes on stakeholders. In the absence of a general account we therefore see many important research questions. For example: What resources might be developed to help practitioners understand how to augment their state spaces or modify their reward functions? How might the research community contribute to making this process more systematic, such as via development of checklists or other design processes? How can developers anticipate what kinds of ethical scenarios must be delineated by the model beyond what is necessary for the task? By what processes can we reliably uncover and anticipate such scenarios---without risking stakeholders---given that many of them are challenging to uncover without deploying a system? 

\textbf{6.4) As with predictive systems, the most straightforward interventions come with considerable drawbacks.} One of the simplest ways developers modify the behavior of MDP agents, whether for ethical or efficiency reasons, is by modifying the reward function. We call this intervention reward modification.\footnote{In reinforcement learning systems with very sparse reward functions---reward functions where most states have the same value, usually zero---a similar sounding technique known as ``reward shaping'' is used to add reward signal to states which represent progress towards or away from one of the original, sparse reward signals. The idea is that the agent will learn faster as it has more frequent access to a learning signal. In the reinforcement learning application there is a lot of concern about executing reward shaping in a manner which does not alter the optimal policy one would get if they solved the original MDP using the original, sparse reward function (remember, however, this is not possible since they do not know the transition function). There are some theoretical results regarding how this may be done \cite{ng1999policy}, but they do not apply to our case because we want to \emph{change} a policy for a \emph{known} MDP.} Reward modification has no direct analogue in prediction, but is similar in spirit to tweaking a loss function in an asymmetric way that affects the model's penalty for incorrect labels on a subset of cases. The important similarities are that the intervention is local, targeting a specific behavior, that the outcome has no formal guarantees since it is unknown how the optimization problem will be re-solved given the new loss function or reward function, and that these interventions require a significant level of expertise since the developer needs to understand how a given change affects some intermediate computation which ultimately affects behavior. Thus, reward modification is necessarily non-methodical. There is no underlying theory that describes how to specify reward functions in order to generate a particular policy or behavior for an arbitrary MDP.

While this intervention is often the easiest, it is also the least effective. Not only is the control over the resultant policy indirect, but this method also leaves room for many tacit normative assumptions. In particular, it allows developers to make implicit comparisons between different types of outcomes due to the reward function mapping all possible outcomes onto the same `unit'. As the agent maximizes expected cumulative reward, it inherently balances avoiding negative reward states and visiting positive reward states based on their respective reward values and the likelihood of reaching those states. Thus, there is always a future amount of positive reward for which the agent will accept experiencing a negative reward in the short term, no matter what real-world scenario that negative reward represents. This is one reason that decision-making model design is so difficult, and this problem is no simpler when modifying reward functions for ethical reasons. That said, in practice this is still the most popular method for modifying agent behavior.

\textbf{6.5) Behavioral constraints on decision-making systems have several benefits.} Beyond gathering more data, expanding decision-making models, and modifying loss functions or reward functions, there are more principled ways to control the behavior of decision-making systems. Generally, these methods constrain the optimization processes involved in determining behavior, and the similarities between techniques devised to produce ``fair'' and ``ethical'' behavior are remarkable. Attempts to train fair predictive models have used constraints \cite{celis2019classification}, regularization \cite{burke2018balanced,kamishima2018recommendation}, and causal and counterfactual analysis \cite{black2020fliptest,black2021leave,coston2020counterfactual}. These techniques essentially constrain the space of possible \emph{functions} that the model can learn. In ethical decision making we typically constrain the space of \emph{policies} using domain-specific hand-coded rules \cite{winfield2014towards,shim2017intervening} or constraints \cite{vanderelst2018architecture,kasenberg2018norm,svegliato2021ethically,nashed2021ethically}. MDP agents that use constraints still compute a policy that maximizes cumulative expected reward, but do so subject to some constraints on, for instance, how often certain state-action pairs can occur. These pairs can be enumerated explicitly or identified via an abstract rule---in the ethical decision-making case, these are rules for acting. This type of MDP is called a constrained MDP (CMDP). However, generating the right constraints is difficult, and is comparable in difficulty to choosing a definition of fairness. There are no clearly best options, and the right choice in terms of satisfying as many stakeholders as possible is often context and deployment specific.

Constraint-based methods are more difficult to design and program but have several advantages. First, this is the only method that guarantees instance-level behavior, although constraints may also be defined in terms of aggregate or expected behavior such that individual decisions may not have guarantees. Second, these methods allow more direct, expressive behavior specification. Instead of changing the reward or loss function, or training using an adversarial agent~\cite{wadsworth2018achieving,zhang2021towards}, we can encode precisely how the agent ought or ought not to behave. Limited only by the expressiveness of the model, the act of writing constraints also surfaces many normative assumptions explicitly. Third, these methods offer substantially greater potential for generalization. Constraints may be formulated in abstract terms, such as false negative rates or the probability of violating a norm, allowing their application to many different decision-making problems. Fourth, although mathematically more complex, these interventions often operate at a level of abstraction that can be communicated to non-experts. This is an important benefit since it allows a greater variety of expertise to be consulted in a given application. While theoretically such constraints have many advantages, these methods are not frequently deployed due to their complexity.\looseness=-1

\textbf{6.6) Non-mathematical and auxiliary interventions, such as human-in-the-loop solutions and explainability, are under-studied in the ethical decision-making context.} Increasing the ability of users or auditors to understand, interact, or correct automated decision-making systems is likely to increase the effectiveness of many existing interventions and perhaps lead to new methods altogether. Interpretable and explainable AI systems are of course large fields in their own right; however, there is a relative lack of research on explainable sequential decision making compared to predictive systems. Not only is there still foundational algorithmic work to be done, but there are also open conceptual questions such as how ideas of actionable recourse \cite{ustun2019actionable,barocas2020hidden} or cross-examination \cite{abebe2022adversarial} might be applied to this setting. Similarly, human-in-the-loop systems have been proposed and studied in the sequential decision-making literature \cite{lam2014pomdp,wray2016hierarchical,feng2016synthesis}, but outside of military research~\cite{brutzman2018ethical,michael2019ethical,swett2021designing,amant2021responsible}, rarely if ever as problems with explicit ethical consequences.

\textbf{6.7) Sequential decision-making models are not generally developed with engagement from the full spectrum of stakeholders.} The opportunity that decision-making models and explicit constraints allow for leveraging expert knowledge cannot be understated. However, current practices in academia and industry do not take full advantage of these benefits in part because they lack exposure to, and knowledge of, qualitative or participatory processes~\cite{martin2020participatory,liao2019enabling,madaio2020co}. By contrast, disciplines such as human-computer interaction have well-developed approaches for engaging with stakeholders, ranging from user-centered design practices \cite{hayes2014knowing} to participatory approaches, where stakeholders work with researchers in a process of collective inquiry \cite{vredenburg2002survey}, or where stakeholders participate in system design and development processes~\cite{paml}.\looseness=-1 

\vspace{-2mm}
\section{Discussion and Conclusion}

This paper explores how the algorithmic fairness and ethical decision-making communities may benefit from knowledge, tools, and practices emerging from one field or the other. In one direction, methods for sequential decision making and modes of analysis from ethical decision making have the potential to advance fairness research given recent calls to examine feedback effects of systems on stakeholder welfare---two themes that have been researched extensively by these communities. In the other, the widespread deployment of sequential decision-making systems and the domains in which they operate (e.g., autonomous driving, power grid management) makes urgent analyses of these systems---analyses which the fairness literature is already undertaking for predictive systems. Here, we imagine analyses of the sequential decision-making model and the processes by which models are designed; the outcomes of decision-making systems in terms of which stakeholders are harmed, particularly the ways in which outcomes might reproduce existing patterns of injustice; and how choices regarding the design of decision-making models give rise to particular outcomes. Alongside these analyses, what processes and resources might we develop to help anticipate the outcomes of a given model and policy, and support safe iterative model development, without incurring too much of the risk inherent to deployment?\footnote{\citet{bird2016exploring} raise a similar question about the risks of autonomous experimentation.}\looseness=-1

Nevertheless, decision-making models are in many ways fundamentally different from predictive models, and their reasoning capabilities, design, and deployment will make realizing these goals difficult. We have illustrated that the interventions---conceptualizations of normative concerns and their accompanying measurements and mitigations---that have entered best practice from the fairness literature may not be applicable to sequential decision making. Moreover, addressing many questions about the design processes, modification, and outcomes of these systems would be prohibitively expensive and likely risky to stakeholders, meaning that such work is likely to be disincentivized in the private sector. Complicating efforts, decision-making systems often operate below the awareness of the public and many regulatory bodies, because they do not tend to make decisions that directly affect individual people. Practical efforts to realize these efforts will require a realistic account of what sequential decision-making systems look like, and of how well our assumptions about what it takes to make fair, transparent, or accountable predictive systems serve us in this different setting.\looseness-1

\bibliographystyle{ACM-Reference-Format}
\bibliography{bibliography}

\end{document}